\documentclass[aps,prl,reprint]{revtex4-2}
\usepackage{amsmath}
\usepackage{amssymb}
\usepackage{graphicx}  
\usepackage{bm}  
\usepackage{xcolor}
\usepackage{printlen}
\usepackage{subcaption}
\usepackage{bbold}
\usepackage{enumitem}
\usepackage{mathtools}
\usepackage{float}
\usepackage{caption}
\usepackage{ragged2e}
\usepackage[
colorlinks,  
linkcolor=blue,  
urlcolor=blue,  
citecolor=blue  
]{hyperref}
\usepackage{url}
\usepackage{float}

\newcommand{\abs}[1]{\left\vert #1 \right\vert}

\newcommand{\bra}[1]{\left\langle #1 \right|}
\newcommand{\ket}[1]{\left| #1 \right\rangle}
\newcommand{\ketbra}[2]{\ket{#1}\!\bra{#2}}
\DeclareMathOperator{\Tr}{Tr}
\DeclareMathOperator{\CD}{CD}

\begin{document}
\title{Cavity cooling using ultrafast electrons}

\author{D. E. Maison}
\altaffiliation{These authors contributed equally to this work}
\author{L. Stettiner}
\altaffiliation{These authors contributed equally to this work}
\author{S. Even-Haim}
\author{A. Gorlach}
\author{I. Kaminer}
\email{kaminer@technion.ac.il}
\affiliation{Technion -- Israel Institute of Technology, Haifa 3200003, Israel}

\begin{abstract}
    We propose a method to cool a thermal photonic state in a cavity by passing electrons through it. Electrons are coherently split into two paths, with one path traversing the cavity, becoming entangled with its photonic state.
    A sequence of such entanglement interactions can achieve cooling of the cavity: e.g., a twofold reduction in thermal photon number with a 25\% post-selection probability. This “which-path”-based approach extends to other qubit-oscillator systems, such as phonons in crystals or optomechanical resonators, offering a general framework for quantum oscillator cooling.
\end{abstract}

\maketitle

\section{Introduction}\label{sec:introduction}
The interaction between electromagnetic fields and electrons is a cornerstone of both theoretical and experimental electrodynamics. Since the explanation of the photoelectric effect a century ago, this field has continued to captivate researchers, particularly in the study of optical near-field interactions with free electrons \cite{barwick:2009, garcia:2010, park:2010, feist:2015, wang:2020, priebe:2017, kfir:2020, dahan:2020}. The interaction of optical near-fields with free electrons offers a powerful tool for probing the quantum properties of light \cite{cahill:1969, Kfir:2019, dahan:2021, gorlach:2024b}. For instance, Ref. \cite{dahan:2023} suggests that energy-shaped free electrons in an optical cavity can generate non-classical light states like cat states \cite{scully:1997, Cochrane:1999} and Gottesman-Kitaev-Preskill (GKP) states \cite{Gottesman:2001}, which are crucial for quantum error correction \cite{terhal:2020, grimsmo:2021, hastrup:2022, deNeeve:2022}.

Such investigations are now part of the field of free-electron quantum optics \cite{Kfir:2019, diGiulio:2019, henke:2021, dahan:2023, ruimy:2025}: facilitating the quantum electron-photon and electron-electron interaction for novel spectroscopy and microscopy \cite{ruimy:2023, gorlach:2024b, ruimy:2024}, for quantum light sources \cite{diGiulio:2022, dahan:2023}, and as a resource for quantum information processing \cite{Baranes:2023}. In all of these, the electrons act as unique flying qubits, performing strong ultrafast interactions.

For these prospects to materialize, the current bottleneck is achieving sufficiently strong electron-photon interactions \cite{adiv:2023, Huang:2023, xie:2025, bezard:2024, karnieli:2024, zhao:2025}. 
The leading schemes involve photonic cavity designs \cite{wang:2020, feist:2022, bezard:2024} at various optical frequencies \cite{yan:2023}. 
Especially at lower frequencies, where most established electron technologies operate, the quantum properties of the interaction are sensitive to the temperature of the cavity. 
The most desired interactions, such as creating electron-photon entanglement \cite{Kfir:2019}, necessitate cooling the cavity to bring its photonic state close to the vacuum state.

More generally, for many quantum technologies, thermal decoherence at finite temperatures poses a significant challenge to precise manipulation of desired quantum states \cite{caldeira:1981, Kim:1992, kim:1992a}.
Thermal photons can hinder high-fidelity quantum operations \cite{brown:2016} by causing a loss of quantum coherence, making it difficult to maintain purity and coherence over time \cite{teh:2020}.

In this work, we propose using the interaction between free electrons and photonic structures to reduce the temperature of their photonic modes.
The key to our scheme is the realization that electron-photon interactions can create a conditional displacement operator on the photonic cavity, which we show can be utilized to cool its state.
The cooling scheme requires a post-selection of each electron state after its interaction, introducing a probabilistic element to the approach.
We analyze both the cooling efficiency and the post-selection probability as functions of interaction parameters and cavity characteristics.

Our free-electron cooling scheme can be explained as a form of many-body ``which-path'' experiment \cite{ruimy:2024}, in which the cavity takes the role of the which-path detector. 
This representation allows applying our proposal to general systems across the fields of continuous-variable quantum information: to cool various qubit-oscillator systems \cite{fluhmann:2019, campagne:2020, eickbusch:2022}.
Relevant oscillator systems include materials with low-energy collective material excitations such as phonons and other forms of polaritons \cite{rivera:2020, kurman:2021}, which could be cooled through their interactions with the free electrons.

\section{Results}\label{sec:theory}
A quantum harmonic oscillator in thermal equilibrium with the environment can be described by the density matrix \cite{landau_3:2013}:
\begin{equation} \label{eq:thermalDensityMatrix}
    \rho_{\bar{n}} = \frac{1}{\bar{n} + 1} \sum\limits_{n=0}^{\infty}\left(\frac{\bar{n}}{\bar{n} + 1}\right)^n \ketbra{n}{n},
\end{equation}
where 
$\bar{n} = \left(\exp\left(\frac{\hbar \omega}{kT}\right) - 1\right)^{-1}$ 
is the average number of thermal photons, with $\omega$ representing the oscillator frequency,
$T$ the temperature, and $\ket{n}$ the $n$-particle Fock state.
Our goal below is to reduce $\bar{n}$, thus lowering the temperature of the system.

A Wigner function \cite{wigner:1932} is an equivalent description of the quantum state, and for the thermal state (\ref{eq:thermalDensityMatrix}) it takes the form \cite{fan:2008}:
\begin{equation} \label{eq:thermalWignerFunction}
    W_{\bar{n}}(x, p) = \frac{1}{\pi \left(2 \bar{n} + 1\right)} \exp\left( -\frac{x^2 + p^2}{2 \bar{n} + 1} \right).
\end{equation}
Here, $x$ and $p$ are real dimensionless quadrature values for position and momentum, respectively.
The thermal Wigner function has a symmetric origin-centered Gaussian form, similar to the vacuum Wigner function $W_{\text{vac}}(x, p) = W_{\bar{n} = 0}(x, p)$, but with a larger variance in both the $x$ and $p$ directions.
The absence of a defined phase for thermal states makes it impossible to decrease the photon number using regular displacement operations.

In the general case, any system coupled to an environment thermalizes,
as described by the Lindblad equation \cite{Saito:1996}, which for a single quantum harmonic oscillator takes the form:
\begin{equation}
	\label{eq:LindbladEquation}
    \begin{split}
        \dot{\rho} = \kappa \left(\bar{n} + 1\right) \left(2a\rho a^\dag - a^\dag a \rho - \rho a^\dag a\right)& \\
        + \kappa \bar{n} \left(2a^\dag\rho a - a a^\dag \rho - \rho a a^\dag\right)&,
    \end{split}
\end{equation}
where $\kappa$ is the cavity dissipation coefficient, and
$a$ and $a^\dag$ are the annihilation and creation operators of the photonic cavity mode, respectively.
This thermalization process brings the system to its thermal equilibrium, i.e., after a sufficiently long duration $t \gg \kappa^{-1}$, the state converges to (\ref{eq:thermalDensityMatrix}), regardless of its initial conditions \cite{Saito:1996, Serafini:2004, leJeannic:2018}.

The cooling process we propose is based on the interaction between free electrons and photonic structures within a transmission electron microscope \cite{park:2010, garcia:2010, Kfir:2019}. 
We specifically consider microwave cavities, where the photonic mode frequency $\omega$ is significantly lower than the energy uncertainty of the free electrons. The scattering matrix that describes the electron-cavity interaction can be represented by the single-mode displacement operator \cite{Kfir:2019, diGiulio:2019, scully:1997}:
\begin{equation} \label{eq:displacement_operator}
    D(g_Q) = \exp\left(g_Q a^\dag - g_Q ^ * a\right),
\end{equation}
where $g_Q$ is the dimensionless electron-cavity interaction parameter, defined as \cite{garcia:2010, park:2010, Kfir:2019}:
\begin{equation}
    g_Q 
    = 
    g_Q(r_\perp)
    = 
    \frac{e}{\hbar\omega} \int dz E_z\left(r_\perp, z\right) \exp{\left(-i\omega z / v\right)}.
\end{equation}
Here, $e$ and $v$ are  the electron charge and velocity, respectively. 
The integration is performed along the electron trajectory ($r_\perp$, $z = v t$), parallel to the $z$ direction, over the electric field component $E_z$, with the field normalized to the energy of a single photon \cite{Kfir:2019}.
Optimizing cavity structures to maximize $|g_Q|$ is an ongoing challenge, with recent works presenting promising bounds \cite{xie:2025, zhao:2025}.

\begin{figure*}
    \centering
    \includegraphics[width=\linewidth]{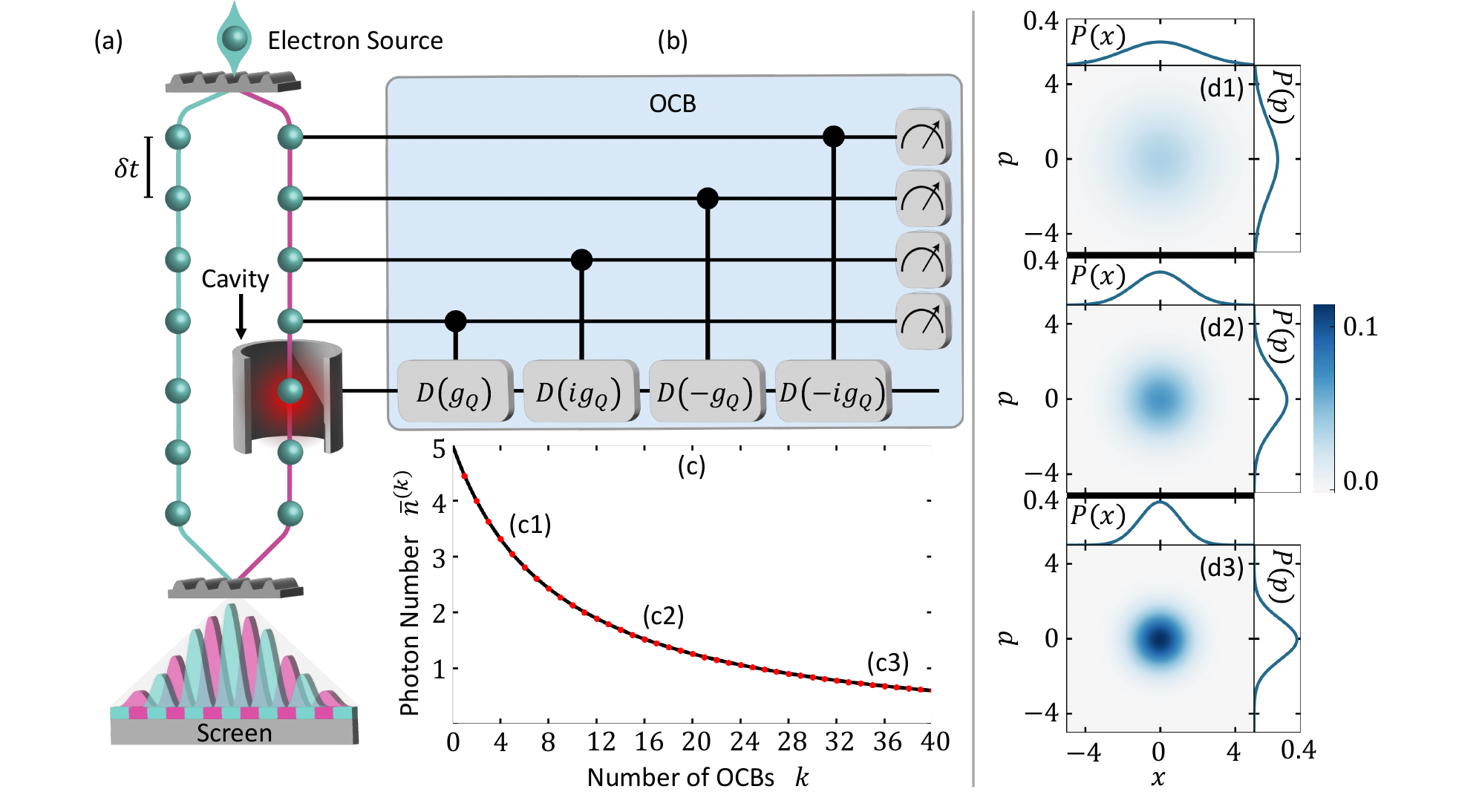}
    \caption{\justifying{
    \textbf{Cavity cooling via free-electron interactions.}
    (a)
    Electrons pass through a beam-splitter
    with time interval $\delta t$ and are split into two paths. 
    The right path corresponds to the $\ket{1}$ state and interacts with the thermal photonic state of a cavity.
    The electron state is measured in the $\ket{\pm}$ basis by observing where it hits the screen relative to the fringes of the interference pattern.
    (b)
    Equivalent quantum circuit representing a single cycle of the cooling process, illustrated as a single Oscillator Cooling Block (OCB).
    (c)
    Number of thermal photons $\bar{n}^{(k)}$ as a function of the number of OCB iterations $k$. The initial state is thermal with $\bar{n}^{(0)}$ = 5, and the electron-cavity interaction parameter is set $g_Q = 0.1$.
    (d1-d3)
    Wigner functions of the states (c1-c3) after 4, 16 and 36 OCB iterations, respectively, as a function of dimensionless position $x$ and momentum $p$ quadratures.
    }}
    \label{fig:f1}
\end{figure*}

We suggest a method of cavity cooling based on electron interference and post-selection. 
Similar to the iconic ``which-path'' experiment, the electrons are first split into two paths, as in split-illumination holography \cite{tanigaki:2012}.
We label the electron's left and right paths as
$\ket{0}$ and $\ket{1}$, respectively, so that before passing through the cavity, the electron is in the state $\ket{+} = (\ket{0} + \ket{1})/\sqrt{2}$.
The photonic structure is positioned along one of these paths (specifically, the right path in this setup), as illustrated in Fig.~\ref{fig:f1}.
The scattering matrix then changes from the displacement operator (\ref{eq:displacement_operator}) to the conditional displacement (CD) operator:
\begin{equation} \label{eq:CD_operator}
    \CD(g_Q) = \ketbra{0}{0} \mathbb{1} + \ketbra{1}{1} D(g_Q).
\end{equation}
Here, the projectors $\ketbra{0}{0}$ and $\ketbra{1}{1}$ 
operate on the electron state,  while $\mathbb 1$ and $D(g_Q)$ act on the photonic state.

The adjoint quantum circuit in Fig. 1\hyperref[fig:f1]{(b)} demonstrates a ``translation'' of the electron-cavity interactions into the language of quantum gates.
This system combines continuous- and discrete-variable quantum information process, with the displacement operators acting on the photonic oscillator state (in the cavity), controlled by the qubits (electrons).
Every four sequential CD operators form together an Oscillator Cooling Block (OCB), which can be repeated multiple times.

Measuring the electron state after the interaction is based on observation of the interference pattern \cite{ruimy:2024}. 
Measuring an electron hit at the location of an even (odd) interference fringe state is $\ket{+}$ ($\ket{-}$).
We post-select on the electron in the $\ket{+}$ state.
Measuring the $\ket{\pm}$ states of the electron is equivalent to applying the following Kraus operator to the photonic mode \cite{evenHaim:2024}:
\begin{equation} \label{eq:Kraus_operator}
    \bra{\pm} \CD(g_Q) \ket{+}=D_\pm(g_Q) 
    = 
    \frac{1}{2}\left(\mathbb{1}  \pm D(g_Q)\right),
\end{equation}
with the expectation value calculated over the electron degree of freedom (its path).

In accordance with the ``sharpen'' technique outlined in Ref. \cite{campagne:2020}, we apply the sequence
$\CD(-ig_Q) \CD(-g_Q) \CD(ig_Q) \CD(g_Q)$ to cool the cavity state.
After each application of the conditional displacement operator, we measure the electron and post-select those in the $\ket{+}$ state.
When heat exchange with the environment can be neglected ($\kappa = 0$), the sequence of OCBs can be represented by the product of Kraus operators (\ref{eq:Kraus_operator}):
\begin{equation} \label{eq:D_OCB}
    D_{\text{OCB}}(g_Q) = D_+(-ig_Q)D_+(-g_Q) D_+(ig_Q) D_+(g_Q).
\end{equation}
This operation transforms the density matrix 
$\rho^{(k+1)}~=~\mathcal{N}(g_Q)~D_{\text{OCB}}(g_Q)~\rho^{(k)} D_{\text{OCB}}^{\dag}(g_Q)$, 
where $\rho^{(k)}$ is the density matrix after $k$ applications of OCB, and $\mathcal{N}(g_Q)$ is a normalization factor. 
We will consider below the case of the initial density matrix being the thermal state (\ref{eq:thermalDensityMatrix}).

If the initial cavity state is thermal (\ref{eq:thermalDensityMatrix}) with the initial photon number $\bar{n}^{(0)}$, 
then for small $\abs{g_Q}$, the state after one OCB application is also close to thermal with 
\begin{equation} \label{eq:new_photon_number}
    \bar{n}^{(1)} = \bar{n}^{(0)} \cdot \left(
            1 - 2\abs{g_Q}^2(\bar{n}^{(0)} + 1)
        \right) + O(\abs{g_Q}^4).
\end{equation}
See Eq. (S8) in Supplemental Material and text below.
The probability to post-select four electrons in the $\ket{+}$ state is
\begin{equation} \label{eq:prob_one_round}
    P^{(1)} = 1 - \abs{g_Q}^2\left(
        2\bar{n}^{(0)} + 1
    \right) + O(\abs{g_Q}^4).
\end{equation}

The choice of $g_Q$ involves a trade-off between the desired temperature reduction and the acceptable probability threshold.
Each OCB iteration can, in principle, select a different value of $g_Q$.
The number of cooling cycles $k$ further controls the degree of cooling versus the post-selection probability.
Figure~\ref{fig:f1}\hyperref[fig:f1]{(c)} shows that the number of thermal photons gradually decreases as more cooling operations are applied, i.e. the temperature decreases.
Figure~\ref{fig:f1}\hyperref[fig:f1]{(d1)}--\hyperref[fig:f1]{(d3)} presents the Wigner functions of the states denoted as \hyperref[fig:f1]{(c1)}--\hyperref[fig:f1]{(c3)}, respectively.

\begin{figure}
    \centering
    \includegraphics[width=\columnwidth,
    clip]{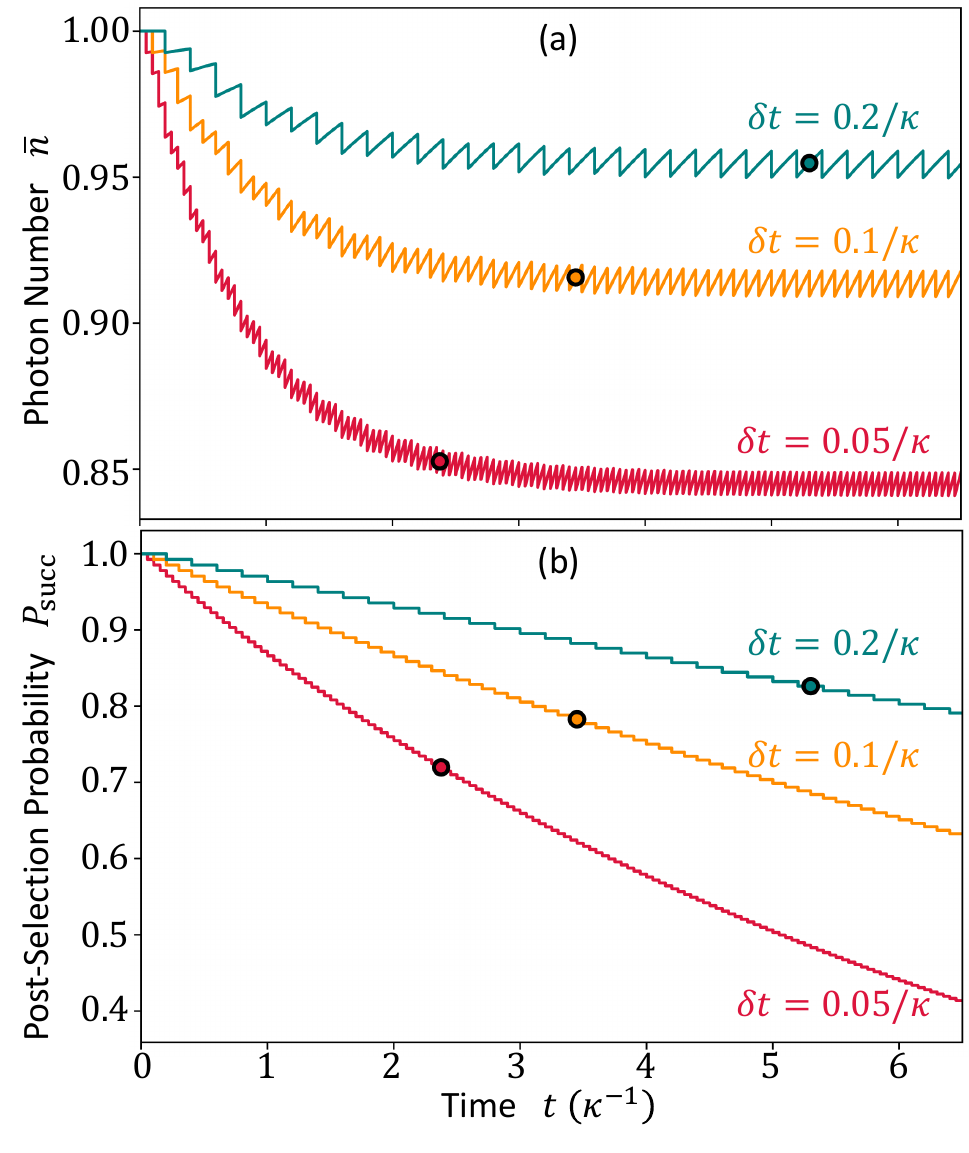}
	\caption{\justifying{
        \textbf{Dynamics of the thermal photon number and post-selection probability.}
    Time evolution of (a) the thermal photon number $\bar{n}(t)$ and (b) the post-selection probability $P_{\text{succ}}(t)$.
    The time $t$ is normalized by the cavity dissipation rate $\kappa$. The discontinuities correspond to applications of CD operations.
    Different curve colors represent varying time intervals $\delta t$ between the consequent electrons.
    Circle markers on both graphs indicate the points where the photon number $\bar{n}$ stabilizes to within a 1\% accuracy (comparing adjacent maxima).
    The interaction constant $g_Q$ is set to 0.1 across all simulations. 
    The initial photon number $\bar{n}^{(0)}= 1$.
    }}
    \label{fig:f2}
\end{figure}

The critical question for implementing the described approach is its efficiency in the case when $\kappa \neq 0$, meaning that the photonic state interacts with the environment between the CD operations.
To address this, we perform numerical modeling of the sequential application of CD operators while the state of the system evolves according to the Lindblad equation (\ref{eq:LindbladEquation}).
We assume that the time interval between the electrons $\delta t$ remains unchanged throughout the process.
We simulate this process numerically \cite{Stettiner:coolingRepo:2024} using the \texttt{QuTiP} package \cite{QuTiP:2012} and present the results in Fig.~\ref{fig:f2}.

Figure~\ref{fig:f2} presents the evolution of the thermal photon number $\bar{n}(t)$ and the successful post-selection probability $P_{\text{succ}}(t)$ 
for having \textit{all} electrons up to the moment $t$ measured in the $\ket{+}$ state.
Different curves represent different time intervals $\delta t$ between electrons.
The thermal photon number $\bar{n}$ exhibits a sawtooth pattern over time.
Each discontinuity corresponds to the application of a CD operator (assuming that the electron-cavity interaction is instantaneous relative to the gradual heating timescale $\kappa^{-1}$).
Between these interactions, the photonic mode gradually heats up as a result of thermalization.
After several cooling cycles, a stable regime is reached, where the cooling effect of the CD operators is balanced by the heating from the environment.
We define the stable regime to begin at the first OCB for which the adjacent maxima differ by less than 1\%, 
indicated by circle markers in Fig.~\ref{fig:f2}\hyperref[fig:f2]{(a)}.
We denote the final stable photon number achieved as $\bar{n}_{\text{f}}$.
As expected, $\bar{n}_{\text{f}}$ increases as $\delta t$ becomes larger, since the photonic mode has more time to restore thermal equilibrium with the bath, before the next electron interaction further cools it.

Figure ~\ref{fig:f2}\hyperref[fig:f2]{(b)} shows the time evolution of the post-selection probability $P_{\text{succ}}(t)$ during the cooling process.
Each application of the CD operator is followed by a post-selection procedure causing the probability to drop.
Between CD operations, no further action is taken in the cavity, so the probability remains constant.
We denote by $P_{\text{f}}$ the probability $P_{\text{succ}}(t)$ at the time that stable $\bar{n}_{\text{f}}$ is achieved, indicated by circle markers in Fig.~\ref{fig:f2}\hyperref[fig:f2]{(b)}.
These probability values $P_{\text{f}}$ are quite sensitive to the accuracy threshold chosen for the photon number (1\% in our case). 
Their values are higher when the accuracy requirement for $\bar{n}$ is less strict.

\begin{figure*}
	\centering
    \includegraphics[width=\linewidth]{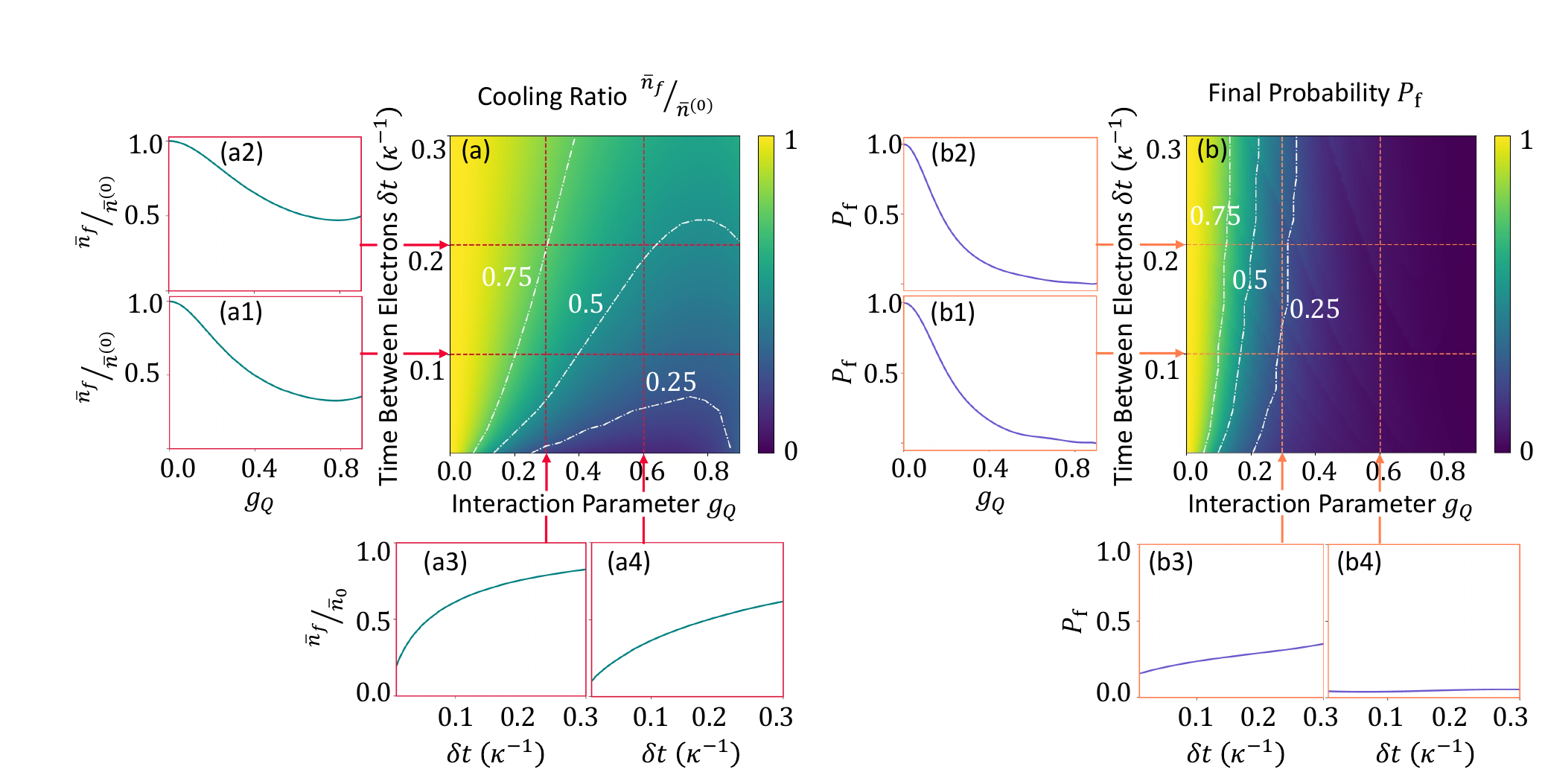}
	\caption{\justifying{
    \textbf{Cooling efficiency and success probability as functions of the interaction parameter $\boldsymbol{g_Q}$ and electron-electron time interval $\boldsymbol{\delta t}$.
    }
    The heatmaps present the ratio of final-to-initial thermal photon numbers $\bar{n}_\textnormal{f} / \bar{n}^{(0)}$ (a) and the post-selection probability $P_{\text{f}}$ (b), both as a function of $g_Q$ (horizontal) and $\delta t$ (vertical).
    White dashed curves highlight contours with specific heatmap values.
    Slices of the heatmap are presented for fixed time intervals between electrons $\delta t=0.1$ (a1, b1) and $\delta t=0.2$ (a2, b2), 
    and for fixed interaction constants $g_Q=0.3$ (a3, b3) and $g_Q=0.6$ (a4, b4).
    }}
    \label{fig:f3}
\end{figure*}

Figure~\ref{fig:f3} provides a detailed analysis of the cooling performance by depicting the final-to-initial photon number ratio $\bar{n}_\text{f} / \bar{n}^{(0)}$ (Fig.~\ref{fig:f3}\hyperref[fig:f3]{(a)}) and the corresponding final post-selection probability $P_{\text{f}}$ (Fig.~\ref{fig:f3}\hyperref[fig:f3]{(b)}) as functions of both $g_Q$ and $\delta t$.
White dashed contour lines on each heatmap indicate specific levels of cooling efficiency (Fig.~\ref{fig:f3}\hyperref[fig:f3]{(a)}) and success probability (Fig.~\ref{fig:f3}\hyperref[fig:f3]{(b)}), with the corresponding values labeled on the plots.
To provide a more detailed view, the auxiliary subplots 
illustrate slices of the heatmaps for fixed time intervals (Fig.~\ref{fig:f3}\hyperref[fig:f3]{(a1,~a2,~b1,~b2)}) and interaction constants (Fig.~\ref{fig:f3}\hyperref[fig:f3]{(a3,~a4,~b3,~b4)}). 
The heatmaps reveal that the most effective cooling is achieved when $g_Q \sim 0.6$ and $\kappa \delta t \ll 0.1$, allowing the photon number to decrease by a factor of ten or more. 
However, this optimal region corresponds to a relatively low success probability of less than 5\%. 
A more moderate cooling effect, such as a twofold reduction in photon number, can be achieved with probability above 25\%. 
As seen from Eq. (\ref{eq:prob_one_round}), the probability of desirable post-selection is higher for lower temperatures. 
Thus, repeating the same procedure will always have a higher success probability for the additional twofold photon number decrease.

\section{Discussion}\label{sec:discussion}

The results underscore a critical trade-off between cooling efficiency and the likelihood of successful post-selection.
Although stronger interactions and shorter electron intervals maximize cooling, they also reduce the probability of success.
Therefore, practical implementation must balance these factors to achieve the desired cooling effect with an acceptable level of success.
For example, in the case of a microwave cavity with frequency $20\ \textrm{GHz}$, the reduction of  average number of thermal photons
from 1 to 0.5 corresponds to a temperature decrease from $1.4 \ \textrm{K}$ to $0.9\ \mathrm{K}$. 
The former can be achieved by pre-cooling of the cavity using liquid helium at low pressure \cite{asztalos:2001}.
The latter is desirable for experiments in free-electron quantum optics \cite{ruimy:2025}, even though still far from the millikelvin level desirable for certain proposals of quantum electron microscopy \cite{okamoto:2012, okamoto:2014, okamoto:2022}.
Stronger cooling is also possible within the suggested protocol, at the price of lower probabilities.

One of the most promising applications of this cooling technique is for improving microwave cavities and detectors, reaching better sensitivity and noise performance by suppressing thermal noise \cite{gao:2008}. This reduction in thermal noise can contribute to ongoing efforts in quantum electron microscopy \cite{okamoto:2012, okamoto:2014, okamoto:2022, koppell:2022} and in dark-matter searches via high-precision measurements \cite{asztalos:2001, du:2018, bertone:2018}.
A direct analogue of our proposal applies to cooling of optical cavities, which can facilitate advancements in quantum optics toward the development of novel quantum light sources. 
Our approach can be generalized to cool certain phononic degrees of freedom of crystalline solids via electron-phonon coupling, which follow similar interaction rules to photons and other polaritons \cite{rivera:2020, kurman:2021}. 
Cooling crystals inside electron microscopes can be key to accessing atomic-resolution imaging of low-temperature phenomena and exotic phase transitions in condensed matter  \cite{feng:2005}.

\section*{acknowledgments}
    We acknowledge Qinghui Yan and Ron Ruimy for valuable discussions and early contributions to this work. 

\bibliographystyle{apsrev}

\newpage

\section*{Supplemental Material}\label{sec:appendix}
\renewcommand{\theequation}{S\arabic{equation}}
\setcounter{equation}{0}
\pagestyle{empty}

Here, we provide the derivation of Eqs. (\ref{eq:new_photon_number}) and (\ref{eq:prob_one_round}). 
We start with the probability $P(+)$ to post-select an electron in the $\ket{+}$ state, given an initial thermal cavity state $\rho_{\bar{n}}$, and calculate it as the trace of the following submatrix:
\begin{multline}
    P(+) = \Tr\left(D_+(g_Q) \rho_{\bar{n}} D_+^{\dag}(g_Q)\right) =
    \\
    \frac{1}{2} + \frac{1}{4}\frac{1}{\bar{n} + 1} \sum\limits_{m=0}^{\infty} \left(\frac{\bar{n}}{\bar{n} + 1}\right)^m \langle m \vert D(g_Q)+D(-g_Q) \vert m \rangle .
\end{multline}
To find the matrix elements, we use results from \cite{cahill:1969}:
\begin{equation}
    \langle m \vert D(g_Q) \vert m \rangle
    = e^{-\frac{\abs{g_Q}^2}{2}} \sum\limits_{m'=0}^{m} \binom{m}{m'} \frac{(-1)^{m'}}{m'!} \abs{g_Q}^{2m'},
\end{equation}
which yields
\begin{equation}
    P(+) = \frac{1}{2} + \frac{e^{-\frac{\abs{g_Q}^2}{2}}}{2(\bar{n} + 1)} \sum\limits_{m,m'} \binom{m}{m'} \! \left(\frac{\bar{n}}{\bar{n} + 1}\right)^m \! \frac{(-\abs{g_Q}^2)^{m'}}{m'!}.
\end{equation}
By rearranging the sums, we find the simplified form:
\begin{equation}
    P(+) = \frac{1}{2} \left (1 + e^{-\abs{g_Q}^2 (\bar{n}+\frac{1}{2})} \right ).
\end{equation}

To cool down the cavity state, we use the following combination of conditional displacement (CD) operators:
\begin{equation*}
    \CD(-ig_Q) \CD(-g_Q) \CD(ig_Q) \CD(g_Q),
\end{equation*}
where the CD operator is defined according to Eq. (\ref{eq:CD_operator}).
The four CDs together form the Oscillator Cooling Block (OCB).
To demonstrate the possibility of cooling down the cavity using the OCB, we consider below the case $\abs{g_Q} \ll 1$.
Up to the first non-trivial order, the corresponding OCB Kraus operator product (\ref{eq:D_OCB}) is expressed as
\begin{equation} \label{eq:D_OCB_approx}
    D_{\text{OCB}}(g_Q) = 1 - \frac{1}{2}\abs{g_Q}^2 \left(2a^\dag a + 1 - i\right) +O\left(\left|g_Q\right|^4\right).
\end{equation}

Applying this operator to the thermal state (\ref{eq:thermalDensityMatrix}) $\rho_{\bar{n}^{(0)}}$ with thermal photon number $\bar{n}^{(0)}$, we get the density matrix after one OCB:
\begin{equation} \label{eq:DM_after_one_OCB}
    \rho^{(1)} 
    \approx 
    \mathcal{N}(g_Q) \sum\limits_{n=0}^{\infty} \left(
        1 - \abs{g_Q}^2 \left(2n + 1\right)
    \right)\left({\rho_{\bar{n}^{(0)}}}\right)_{nn} \ketbra{n}{n},
\end{equation}
where $\left(\rho_{\bar{n}^{(0)}}\right)_{nn}$ is the $n$-th diagonal element of $\rho_{\bar{n}^{(0)}}$.
The normalization factor $\mathcal{N}(g_Q)$ is defined from the identity $\Tr\rho^{(1)} = 1$ and is therefore given by
\begin{equation}
    \mathcal{N}(g_Q) \approx \frac{1}{1 - \abs{g_Q}^2 \left(2\bar{n}^{(0)} + 1\right)}.
\end{equation}
The thermal photon number after one OCB and post-selection is
$\bar{n}^{(1)} = \Tr\left(a^\dag a\rho^{(1)}\right)$, which up to order $\abs{g_Q}^2$ equals
\begin{equation} \label{eq:approx_photon_num_one_round}
    \bar{n}^{(1)} = \bar{n}^{(0)} \cdot \left(
        1 - 2\abs{g_Q}^2\left(\bar{n}^{(0)} + 1\right)
    \right) + 
    O\left(\abs{g_Q}^4\right).
\end{equation}
Within the same level of accuracy, Eq. (\ref{eq:DM_after_one_OCB}) coincides with the expression for thermal state $\rho_{\bar{n}^{(1)}}$ with the photon number $\bar{n}^{(1)}$.
Thus, up to order $\abs{g_Q}^2$, the state remains thermal, with its temperature decrease depending both on the interaction parameter $g_Q$ and on the photon number $\bar{n}^{(0)}$.

The probability of post-selecting all four electrons in the $\ket{+}$ state is given by
\begin{equation} \label{eq:approx_prob_one_round}
    \begin{split}
        P^{(1)} &= \Tr\left(
            D_{\text{OCB}}(g_Q)\rho_{\bar{n}^{(0)}} D_{\text{OCB}}^\dag(g_Q)
        \right) 
        = 
        \frac{1}{\mathcal{N}(g_Q)} \\
        &\approx 1 - \abs{g_Q}^2 \left(2\bar{n}^{(0)} + 1\right)
    \end{split}
\end{equation}

After $k$ repetitions of the cooling procedure, the photon number and probability can be obtained by multiple recursive application of Eqs.~(\ref{eq:approx_photon_num_one_round}, \ref{eq:approx_prob_one_round}).
To gain intuition, we examine the limit of $k \ll \abs{g_Q}^{-2}$, for which a simplified analytical result arises since the temperature does not decrease significantly during the cooling. In this case, one can neglect the variation of the temperature factor inside the brackets, and obtain:
\begin{equation} \label{eq:approx_photon_num_several_rounds}
    \begin{split}
    \bar{n}^{(k)} &\approx
    \bar{n}^{(0)} \cdot \left(
        1 - 2\abs{g_Q}^2\left(\bar{n}^{(0)} + 1\right)
    \right)^k \\
    &\approx
    \bar{n}^{(0)} \cdot \left(
        1 - 2k\abs{g_Q}^2\left(\bar{n}^{(0)} + 1\right)
    \right),
    \end{split}
\end{equation}

\begin{equation} \label{eq:approx_prob_several_rounds}
    \begin{split}
    P^{(k)} &\approx 
    \left( 
        1 - \abs{g_Q}^2\left(2\bar{n}^{(0)} + 1\right)
    \right)^k \\
    &\approx
    1 - k\abs{g_Q}^2\left(2\bar{n}^{(0)} + 1\right).
    \end{split}
\end{equation}

{\color{white} \ }  

\end{document}